\documentclass[aps,pre,amsmath,twocolumn,superscriptaddress]{revtex4-1}
\usepackage{graphicx}
\usepackage{color}
\usepackage{bm}
\usepackage{braket}
\usepackage{amsmath,amssymb}

\begin{document}
\title{Structure and lattice thermal conductivity of grain boundaries in silicon by using machine learning potential and molecular dynamics}
\author{Susumu \surname{Fujii}}
\email{susumu.fujii@mat.eng.osaka-u.ac.jp}
\affiliation{Nanostructures Research Laboratory, Japan Fine Ceramics Center, Nagoya 456-8587, Japan}
\affiliation{Division of Materials and Manufucturing Science, Osaka University, Osaka 565-0871, Japan}
\author{Atsuto \surname{Seko}}
\email{seko@cms.mtl.kyoto-u.ac.jp}
\affiliation{Department of Materials Science and Engineering, Kyoto University, Kyoto 606-8501, Japan}

\date{\today}

\begin{abstract}
In silicon, lattice thermal conductivity plays an important role in a wide range of applications such as thermoelectric and microelectronic devices.
Grain boundaries (GBs) in polycrystalline silicon can significantly reduce lattice thermal conductivity, but the impact of GB atomic structures on it remains to be elucidated.
This study demonstrates accurate predictions of the GB structures, GB energies, and GB phonon properties in silicon using machine learning potentials (MLPs).
The results indicate that the MLPs enable robust GB structure searches owing to the fact that the MLPs were developed from a training dataset covering a wide variety of structures.
We also investigate lattice thermal conduction at four GB atomic structures using large-scale perturbed molecular dynamics and phonon wave-packet simulations.
The comparison of these results indicates that the GB structure dependence of thermal conductivity stems from anharmonic vibrations at GBs rather than from the phonon transmission behavior at GBs.
The advantages of the MLPs compared with a typical empirical potential of silicon are also thoroughly investigated.
\end{abstract}

\maketitle

\section{Introduction}

Lattice thermal conductivity plays an essential role in a wide variety of applications in silicon 
\cite{
Bux2009,
Yusufu2014,
Nakamura2015,
Sakata2015,
Oyake2018,
Uma2001,
Schelling2005,
Pop2010,
Moore2014,
Liebchen2016}.
For example, low thermal conductivity provides an efficient energy conversion in thermoelectrics, whereas high thermal conductivity is desirable for heat dissipation in micro- or nanoelectronic devices.
In the former case, polycrystalline silicon with a grain size of micro- or nanometer order has been synthesized to introduce a large number density of grain boundaries (GBs), which reduce the thermal conductivity caused by the interfacial thermal resistance \cite{Bux2009,Yusufu2014,Nakamura2015,Sakata2015,Oyake2018}.
In contrast, in the latter case, the thermal resistance of interfaces limits heat dissipation and exacerbates thermal management problems \cite{Uma2001, Schelling2005, Pop2010, Moore2014, Liebchen2016}, although monocrystalline silicon is highly thermally conductive (156 W/mK at 300 K) \cite{Glassbrenner1964}.
Thus, there has been an increasing desire to understand the mechanism of GB thermal conduction in silicon quantitatively, both in terms of heat insulation and dissipation.

Experimental thermal conductivity measurements in polycrystalline silicon demonstrated a significant impact of GBs on thermal conductivity \cite{Wang2011a,Jugdersuren2017}.
They revealed that the thermal conductivity of polycrystalline silicon with a grain size of 9.7 nm is more than 95\% lower than that of monocrystalline silicon \cite{Jugdersuren2017}.
However, the impact of individual GBs and their atomistic structures has only been evaluated experimentally in a few studies owing to the technical difficulties involved \cite{Xu2018}.
Instead, computational techniques such as molecular dynamics (MD) have revealed the microscopic features of GB thermal conduction and its relationship with GB atomistic structures 
\cite{
Bodapati2006,
Kimmer2007,
Aubry2008,
Ju2012,
Ju2013,
Bohrer2017,
Hickman2020,
Chernatynskiy2016,
Fujii2019,
Fujii2020}.
For example, in previous studies, the transmission and reflection behaviors of acoustic phonons in several twist GBs were investigated by the phonon wave packet method \cite{Kimmer2007,Ju2013}, and the lattice thermal conductivity in the $\langle 001 \rangle$ symmetric tilt GBs was systematically calculated using the nonequilibrium MD method, the so-called direct method \cite{Hickman2020}.
These studies clarified the frequency dependence of phonon transmission and the dependence of thermal conductivity on the GB misorientation for a given set of silicon GB structures.

Such large-scale computational simulations have employed empirical interatomic potentials of the Stillinger--Weber (SW) \cite{Stillinger1985,Watanabe1999} and Tersoff \cite{Tersoff1988,Tersoff1989,Munetoh2007} potentials, developed by reproducing only a limited number of physical properties of prototype structures, including the diamond-type structure.
Therefore, these potentials should be less transferable to complex atomic structures found in GBs as follows.
Firstly, the most stable GB atomic structures depend on the function form and the inferred parameters of an empirical potential, as extensively investigated for the $\langle110\rangle$ symmetric tilt GBs \cite{Wang2019}.
Additionally, empirical potentials highly underestimate or overestimate the lattice thermal conductivity of silicon in the bulk form \cite{Howell2012}, which implies that the thermal behavior predicted using empirical potentials is artificial.
Thus, the applicability of these empirical potentials in GB thermal conduction still needs to be examined.

In contrast to the use of empirical potentials, density functional theory (DFT) calculations can reproduce the experimental lattice thermal conductivity of monocrystalline silicon \cite{Broido2007,Esfarjani2011,Rohskopf2017} and atomic structures of several GBs observed by electron microscopy \cite{Kohyama1988,Maiti1996,Shi2010,Sakaguchi2007}.
Unfortunately, the computational cost of applying DFT calculations to GB models is high, which means that it is challenging to perform GB thermal conduction analyses.

Machine learning potentials (MLPs) have been increasingly developed from extensive datasets generated by DFT calculations to simultaneously achieve high transferability and low computational cost.
Therefore, MLPs are becoming valuable tools for performing crystal structure optimizations and accurate large-scale atomistic simulations, which are prohibitively expensive when performed by DFT calculations
\cite{
Lorenz2004210,
behler2007generalized,
bartok2010gaussian,
behler2011atom,
han2017deep,
258c531ae5de4f5699e2eec2de51c84f,
PhysRevB.96.014112,
PhysRevB.90.104108,
PhysRevX.8.041048,
PhysRevLett.114.096405,
PhysRevB.95.214302,
PhysRevB.90.024101,
PhysRevB.92.054113,
PhysRevMaterials.1.063801,
Thompson2015316,
wood2018extending,
PhysRevMaterials.1.043603,
PhysRevB.98.094104,
doi-10.1137-15M1054183,
PhysRevLett.120.156001,
podryabinkin2018accelerating,
GUBAEV2019148,
doi:10.1063/1.5126336}.
Several MLPs have been proposed for silicon using the Gaussian process and artificial neural networks \cite{PhysRevX.8.041048,YokoiPRM2020,doi:10.1063/5.0013826,deringer2021origins}.
Moreover, MLPs have been recently applied to predicting GB properties \cite{PhysRevMaterials.4.123607,YokoiPRM2020,ZHENG202040}.
One of the authors also developed a set of Pareto optimal MLPs for silicon, available in the \textsc{Machine Learning Potential Repository} \cite{MachineLearningPotentialRepositoryArxiv,MachineLearningPotentialRepository}.
These MLPs describe the potential energy using polynomial models of systematic polynomial invariants for the O(3) group.

In this study, we systematically investigate stable GB structures of silicon and analyze their lattice thermal conduction behavior using an appropriate MLP from the set of Pareto optimal MLPs.
The lattice thermal conduction behavior is estimated using a combination of the appropriate MLP, a rigid-body translation approach, perturbed MD, and the phonon wave packet method.
We also demonstrate the transferability of the MLP to the prediction of GB properties throughout this study.

In Sec. II, we introduce the present computational procedure for developing the Pareto optimal MLPs and selecting an appropriate MLP for predicting GB properties.
In Sec. III, we show the prediction of the stable GB structure, GB excessive energy, and phonon properties at GBs.
In Sec. IV, the thermal conduction behavior at $\Sigma5(310)$ and $\Sigma3(112)$ GBs is examined using the perturbed MD and phonon wave packet method.
The study is summarized in Sec. V.

\section{Machine learning potentials}
\subsection{Development of MLPs}
In this study, we employ MLPs for silicon selected from the set of Pareto optimal MLPs with different trade-offs between accuracy and computational efficiency in the \textsc{Machine Learning Potential Repository} \cite{MachineLearningPotentialRepositoryArxiv,MachineLearningPotentialRepository} developed by one of the authors.
The Pareto optimal MLPs were developed using a dataset generated from 86 structure generators, i.e., unique prototype structures composed of single elements with zero oxidation state included in the Inorganic Crystal Structure Database (ICSD) \cite{bergerhoff1987crystal}.
The dataset comprises 10,000 structures, each of which was generated by introducing random lattice expansion, random lattice distortion, and random atomic displacements into a supercell of the equilibrium structure for one of the structure generators \cite{PhysRevB.99.214108}. 
DFT calculations were performed for all structures in the dataset using the plane-wave-basis projector augmented wave method \cite{PAW1} within the Perdew--Burke--Ernzerhof exchange-correlation functional \cite{GGA:PBE96} as implemented in the \textsc{vasp} code \cite{VASP1,VASP2,PAW2}. 
The cutoff energy was set to 300 eV.
The total energies converged to less than 10$^{-3}$ meV/supercell.
The equilibrium atomic positions and lattice constants for the structure generators were optimized until the residual forces are less than 10$^{-2}$ eV/\AA.
The dataset was then randomly divided into a training dataset and a test dataset.
Note that both the datasets contain no structures generated from GB models.

Polynomial models of polynomial invariants, including high-order ones \cite{PhysRevB.99.214108,PhysRevB.102.174104}, were then systematically applied to develop MLPs.
Because the accuracy and computational efficiency of the MLPs strongly depend on several input parameters, such as the cutoff radius, the type of structural feature, the type of potential energy model, the number of radial functions, the truncation of the polynomial invariants, and the polynomial order in the potential energy model, a systematic grid search was performed to find their optimal values.
As a result of the grid search, a set of 38 Pareto optimal MLPs was obtained from a total of 1001 potential energy models.
Such MLPs represented by polynomial models of polynomial invariants has been found to accurately predict the GB energy in metallic systems \cite{PhysRevMaterials.4.123607}.
Regression coefficients of a potential energy model were estimated by linear ridge regression, simultaneously using the total energies, the forces acting on atoms, and the stress tensors computed by the DFT calculations for structures in the training dataset.
Further details on the regression can be found in Ref. \onlinecite{PhysRevB.99.214108}.

\begin{figure}[tbp]
\includegraphics[clip,width=\linewidth]{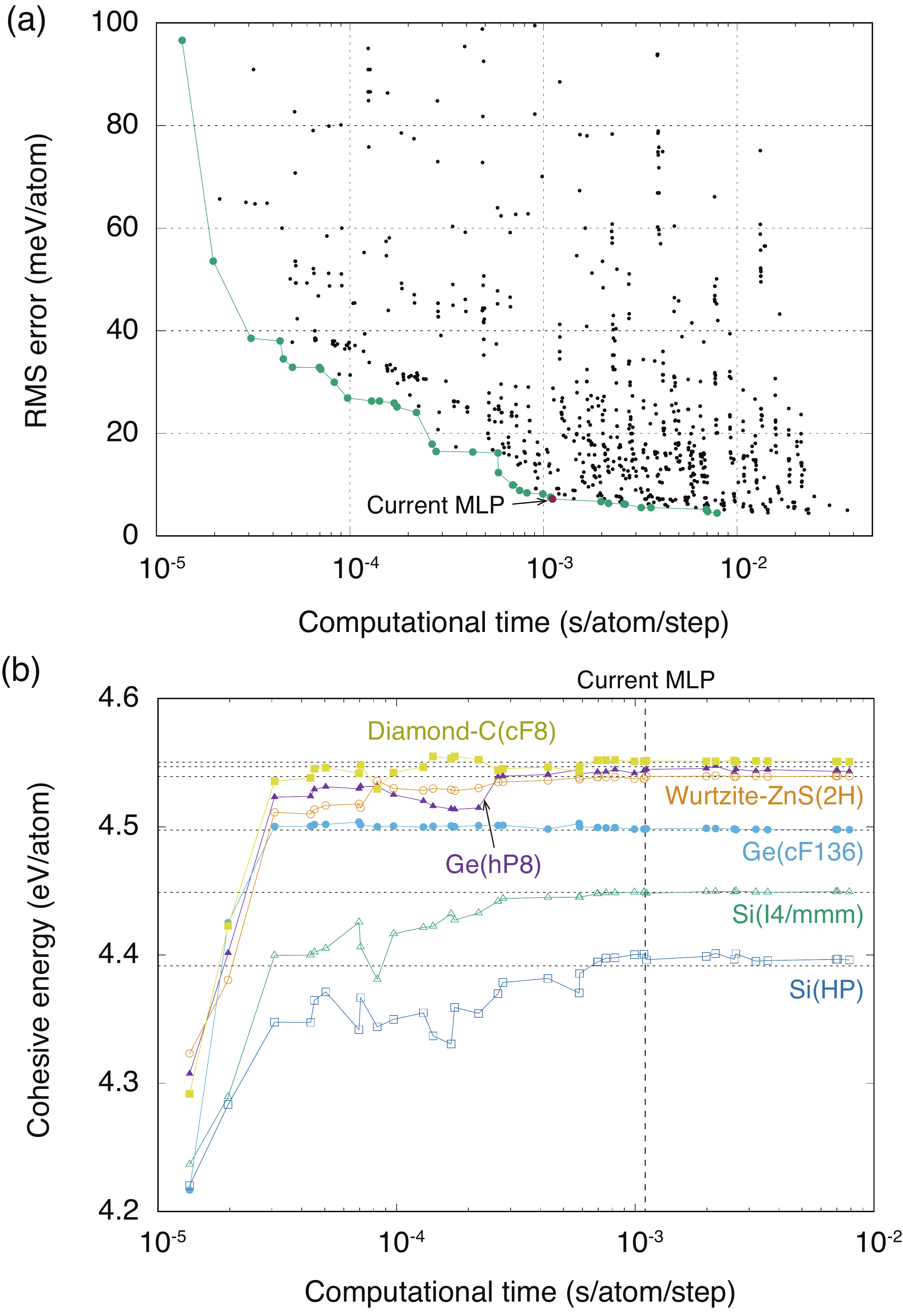}
\caption{
(a) Distribution of MLPs for silicon.
The green and purple closed circles show Pareto optimal points of the distribution with different trade-offs between accuracy and computational efficiency.
The current MLP is shown by purple closed circles.
The computational time indicates the elapsed time for a single-point calculation normalized by the number of atoms.
The elapsed time is measured using a single core of Intel\textregistered\ Xeon\textregistered\ E5-2695 v4 (2.10 GHz).
(b) Cohesive energies predicted using the Pareto optimal MLPs for the first six structures in the descending order of the cohesive energy. 
They are Diamond-C(cF8)-, Ge(hP8)-, Wurtzite-ZnS(2H)-, Ge(cF136)-, Si(I4/mmm)-, and Si(HP)-type structures, which are denoted by the ICSD notation.
Their DFT-calculated values are also shown by the dotted lines.
}
\label{SF-AS:Fig-pareto}
\end{figure}

Figure \ref{SF-AS:Fig-pareto}(a) shows the distribution of the MLPs obtained from the grid search. 
Here, the root mean square (RMS) error for the test dataset is used as an estimator of the accuracy of the MLPs.
The computational time associated with the model complexity of an MLP corresponds to the elapsed time normalized by the number of atoms for a single-point calculation of the energy, forces, and stress tensors.
As can be seen in Fig. \ref{SF-AS:Fig-pareto}(a), the accuracy and computational efficiency of the MLPs are conflicting properties. 
Therefore, the Pareto optimal MLPs with different trade-offs between the accuracy and computational efficiency turn out to be optimal as shown in Fig. \ref{SF-AS:Fig-pareto}(a).

\subsection{Selection of MLP}

\begin{figure*}[tbp]
\includegraphics[clip,width=0.9\linewidth]{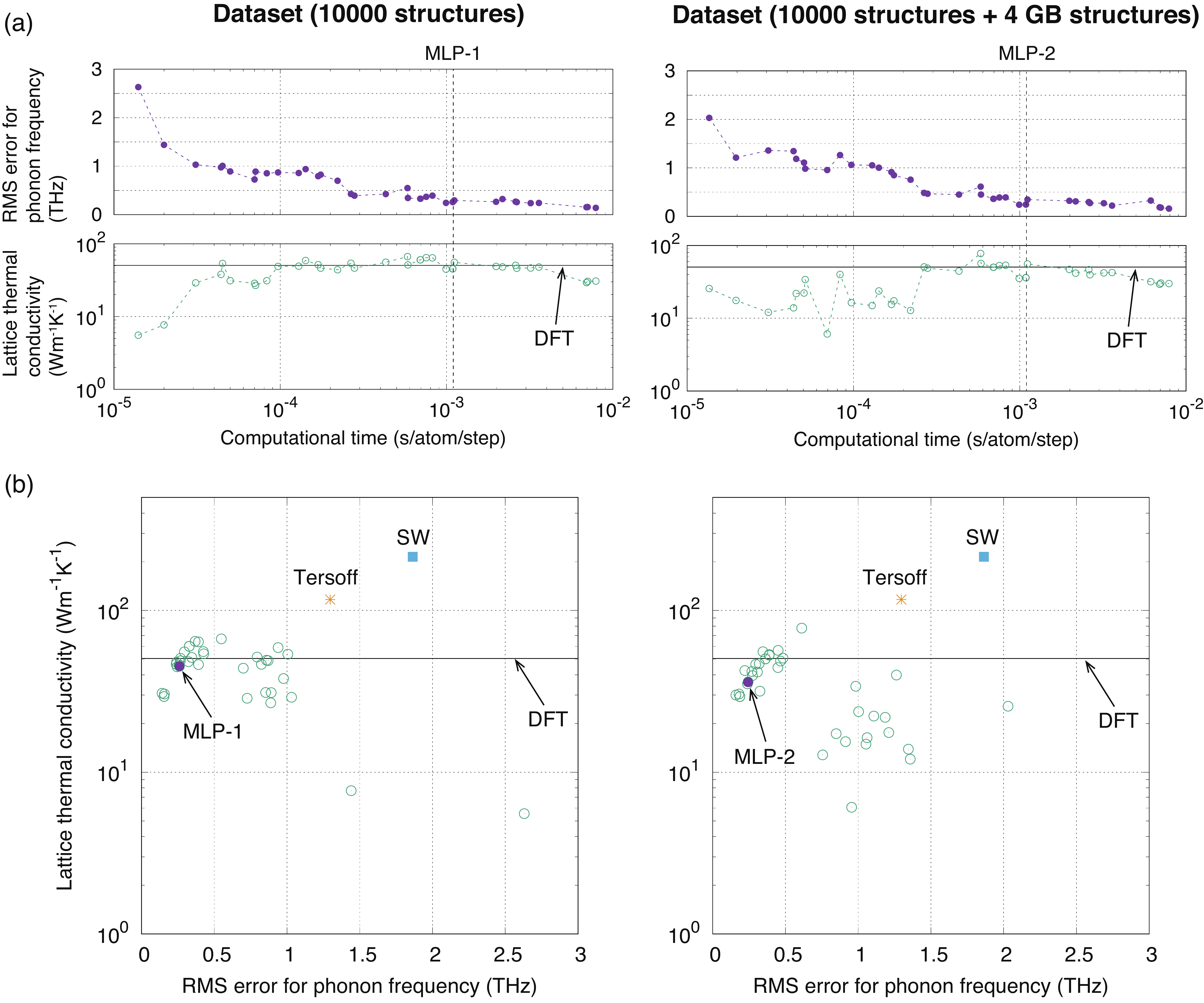}
\caption{
(a) RMS errors of the phonon frequency for the diamond-type structure predicted using the Pareto optimal MLPs. 
The lattice thermal conductivities for the diamond-type structure at 700 K predicted using the Pareto optimal MLPs are also shown.
(b) Dependence of the lattice thermal conductivity on the RMS error of the phonon frequency, replotted from Fig. \ref{SF-AS:Fig-convergence}(a).
The left panels show the properties predicted using the Pareto optimal MLPs developed from the dataset composed of 10,000 structures.
The right panels show the properties predicted using the MLPs estimated from the dataset composed of 10,000 structures and four RBT(MLP-1) structures.
}
\label{SF-AS:Fig-convergence}
\end{figure*}

To choose an appropriate MLP from the Pareto optimal ones, we evaluate the cohesive energy for several structures, the phonon frequencies for the diamond-type structure, and the lattice thermal conductivity for the diamond-type structure at 700 K using the entire set of Pareto optimal MLPs.
We use our implementation of the MLPs (\textsc{lammps-mlip-package} \cite{LammpsMLIPpackage}) that enables us to use the MLPs in the \textsc{lammps} code \cite{Plimpton1995}.
The package is also employed for evaluating GB structures and properties in the following sections.
Figure \ref{SF-AS:Fig-pareto}(b) shows the convergence behavior of the cohesive energy in terms of the computational time of the MLPs for the first six structures in the descending order of the cohesive energy.
Figure \ref{SF-AS:Fig-convergence}(a) also shows the RMS error for the phonon frequency and lattice thermal conductivity for the diamond-type structure predicted using the entire set of Pareto optimal MLPs.
The phonon properties and lattice thermal conductivity are calculated using \textsc{phonopy} \cite{phonopy2} and \textsc{phono3py} \cite{phono3py}, respectively.
As shown in Figs. \ref{SF-AS:Fig-pareto}(b) and \ref{SF-AS:Fig-convergence}(a), the cohesive energy and lattice thermal conductivity converge well to their DFT calculated values.
Moreover, the RMS error for the phonon frequency approaches zero as the model complexity increases.

Figure \ref{SF-AS:Fig-convergence}(b) shows the MLPs and empirical potentials \cite{Stillinger1985,Tersoff1989} with respect to the RMS error for the phonon frequency and lattice thermal conductivity.
The SW and Tersoff potentials show larger prediction errors for the phonon frequency and lattice thermal conductivity.
On the basis of these convergence behaviors of the Pareto optimal MLPs, we employ an MLP with a computational cost of approximately 1 ms/atom/step to predict successive GB properties.
The potential energy model of the current MLP and its predictive power for the other physical properties are summarized in Appendix \ref{SF-AS:Appendix-MLP}.
The current MLP will be hereafter referred to as ``MLP-1''.

\section{Grain boundary properties}
\subsection{Rigid-body translation approach}
We adopt a rigid-body translation (RBT) approach to find the globally optimal atomic configuration for each GB.
An initial atomic configuration is generated by shifting one crystal relative to another crystal by two components parallel to the GB plane ($y$- and $z$-axes) and a component perpendicular to the GB plane  ($x$-axis) \cite{Priester}.
Here, we set the increment of the RBT in the $y$- and $z$-directions to approximately 0.5 \AA\, depending on the GB.
The separation of the two crystals in the $x$-direction is given by a sequence of 0.5, 1.0, 1.5, and 2.0 \AA.
If two atoms are less than half of the bond length in the diamond-type structure of silicon, we replace them with an atom located at the center of the two atoms.
By performing the local geometry optimization for all the initial configurations, we obtain the GB energy surface as a function of the RBTs and the most stable atomic configuration.
Hereafter, the most stable GB structure predicted using the potential $p$ will be referred to as the ``RBT($p$) structure''.

\subsection{GB structures and energies}

\begin{table}[tbp]
\begin{ruledtabular}
\caption{
GB energies for the RBT(MLP-1) structures predicted using the SW potential, Tersoff potential, and MLPs, and by DFT calculations in units of mJ/m$^2$.
MLP-2 corresponds to MLP-1 with a slight modification, which is developed from a training dataset including the DFT energies of the RBT(MLP-1) structures for $\Sigma5(310)$, $\Sigma3(112)$, and $\Sigma21(1\bar54)$, and the HRTEM structure for $\Sigma3(112)$.
The upper four GB structures are used for developing MLP-2, whereas the lower four GB structures are not used for developing MLP-2.
}
\label{SF-AS:Table-gb-energy}
\begin{tabular}{lccccc}
GB & SW & Tersoff & MLP-1 & MLP-2 & DFT \\
\hline
$\Sigma3(112)/[1\bar10]$             & 931  & 803 & 757 & 715 & 674  \\
$\Sigma3(112)/[1\bar10]$\footnote{HRTEM structure \cite{Sakaguchi2007}} 
                                     & 702  & 698 & 758 & 433 & 375  \\
$\Sigma5(310)/[001]$                 & 661  & 629 & 622 & 387 & 333  \\
$\Sigma21(1\bar54)/[111]$            & 811  & 795 & 699 & 575 & 534  \\
\hline
$\Sigma11(113)/[1\bar10]$            & 874  & 881 & 916 & 722 & 676  \\
$\Sigma5(210)/[001]$                 & 673  & 649 & 604 & 409 & 361  \\
$\Sigma9(221)/[1\bar10]$             & 446  & 440 & 462 & 252 & 185  \\
$\Sigma19(331)/[1\bar10]$            & 498  & 496 & 493 & 337 & 285  \\
\hline
RMS error                            & 276  & 253 & 250 &  52 & $-$  \\
\end{tabular}
\end{ruledtabular}

\end{table}

\begin{figure}[tbp]
\includegraphics[clip,width=\linewidth]{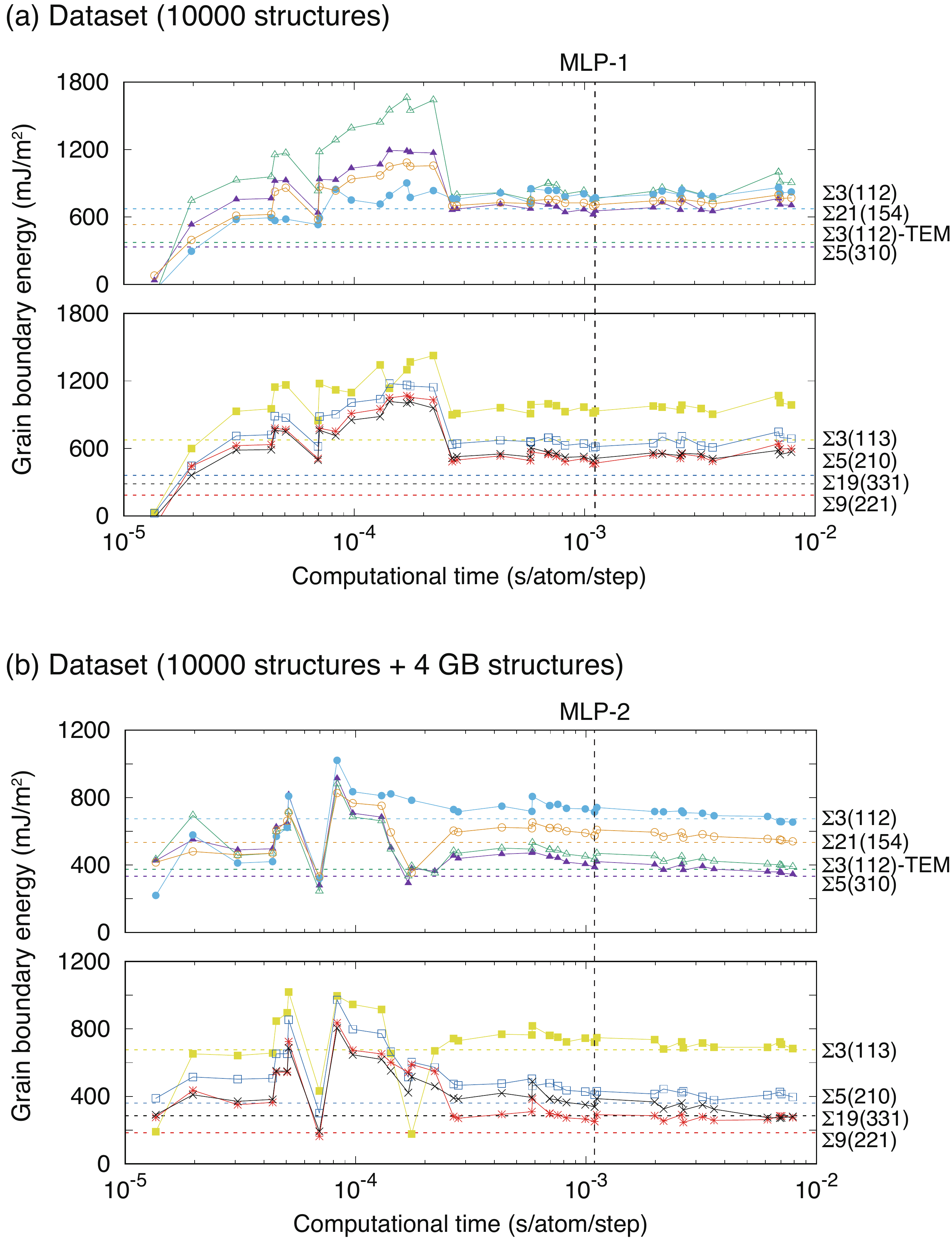}
\caption{
(a) GB energies predicted using the Pareto optimal MLPs, which are estimated from the dataset containing no GB structures.
The DFT-calculated GB energies are also shown by the dotted lines.
(b) GB energies obtained using the Pareto optimal MLPs, which are estimated from the dataset including four GB structures.
}
\label{SF-AS:Fig-gb-energy}
\end{figure}

We search for the stable structures of seven symmetric tilt GBs, i.e., $\Sigma3(112)$, $\Sigma5(310)$, $\Sigma21(1\bar54)$, $\Sigma11(113)$, $\Sigma5(210)$, $\Sigma9(221)$, and $\Sigma19(331)$ GBs, using the RBT approach and MLP-1.
To examine the predictive power of MLP-1 for the GB energy, we also compute the GB energies using the empirical potentials and by DFT calculations.
Table \ref{SF-AS:Table-gb-energy} shows the GB energies of the RBT(MLP-1) structures for the seven GBs predicted using the SW potential, the Tersoff potential, and MLP-1 compared with those predicted by the DFT calculations.
Although MLP-1 and the empirical potentials overestimate the GB energy predicted by the DFT calculations, they almost reconstruct the hierarchy of the GB energy for the seven GBs.
Figure \ref{SF-AS:Fig-gb-energy}(a) shows the GB energy of the RBT(MLP-1) structures for the seven GBs predicted by the entire set of Pareto optimal MLPs.
As well as MLP-1, the other Pareto optimal MLPs overestimate the GB energy, although the GB energy converges well in terms of the model complexity.

In the literature, among the seven GBs, the structures of $\Sigma5(310)$ \cite{Bourret1988}, $\Sigma3(112)$ \cite{Sakaguchi2007}, and $\Sigma9(221)$ GBs \cite{Couillard2013} have been clarified experimentally.
The experimental structures were found to be consistent with the RBT(MLP-1) structures for $\Sigma5(310)$ and $\Sigma9(221)$ GBs by comparing the experimental GB structures with the RBT(MLP-1) structures.
On the other hand, the simple RBT approach does not derive the $\Sigma 3(112)$ structure observed by high-resolution transmission electron microscopy (HRTEM) \cite{Sakaguchi2007}.
To reach the complex HRTEM structure, additional operations such as the removal of atoms at the GB plane \cite{Hickman2017} may be required before performing the local geometry optimization.

To examine the stability of the HRTEM structure for $\Sigma 3(112)$ GB, we manually reconstruct the atomic configuration of the HRTEM structure and evaluate its GB energy using MLP-1 and the empirical potentials, and by DFT calculations.
Table \ref{SF-AS:Table-gb-energy} shows the predicted GB energy of the HRTEM structure.
The GB energy of the HRTEM structure computed by the DFT calculations is 375 mJ/m$^2$, which is lower than that of the RBT(MLP-1) structure, whereas the GB energy of the HRTEM structure predicted using MLP-1 is almost the same as that of the RBT(MLP-1) structure.

As described above, MLP-1 and the other Pareto optimal MLPs in the repository were developed using the training dataset that contains no GB structures.
Therefore, we develop a slightly modified MLP of MLP-1 by including some of the GB structures in the training dataset to improve the predictive power for the GB energy.
MLP-1 and the modified MLP have the same potential energy model, and the modified MLP is estimated from a training dataset that includes the energies of the equilibrium GB structures computed by the DFT calculations.
They are the RBT(MLP-1) structures for $\Sigma5(310)$, $\Sigma3(112)$, and $\Sigma21(1\bar54)$ GBs, and the HRTEM structure for $\Sigma3(112)$ GB.
That is, only four DFT energies for the GB structures are appended to the training dataset.
The regression coefficients of the modified MLP are then estimated by weighted linear ridge regression with large weights for the GB structures.
The modified MLP is hereafter referred to as ``MLP-2''.

Table \ref{SF-AS:Table-gb-energy} shows the GB energies for the RBT(MLP-1) structures predicted using MLP-2.
The training dataset does not contain the last four GB structures in Table \ref{SF-AS:Table-gb-energy}, i.e., $\Sigma11(113)$, $\Sigma5(210)$, $\Sigma9(221)$, and $\Sigma19(331)$ GBs.
Figure \ref{SF-AS:Fig-gb-energy}(b) shows the GB energies for the eight GB structures predicted using MLP-2 and the other Pareto optimal MLPs developed using the training dataset with the GB structures.
The GB energy converges well in terms of model complexity, and the converged GB energies are close to the DFT-calculated GB energies.

\begin{figure}[tbp]
\includegraphics[clip,width=0.8\linewidth]{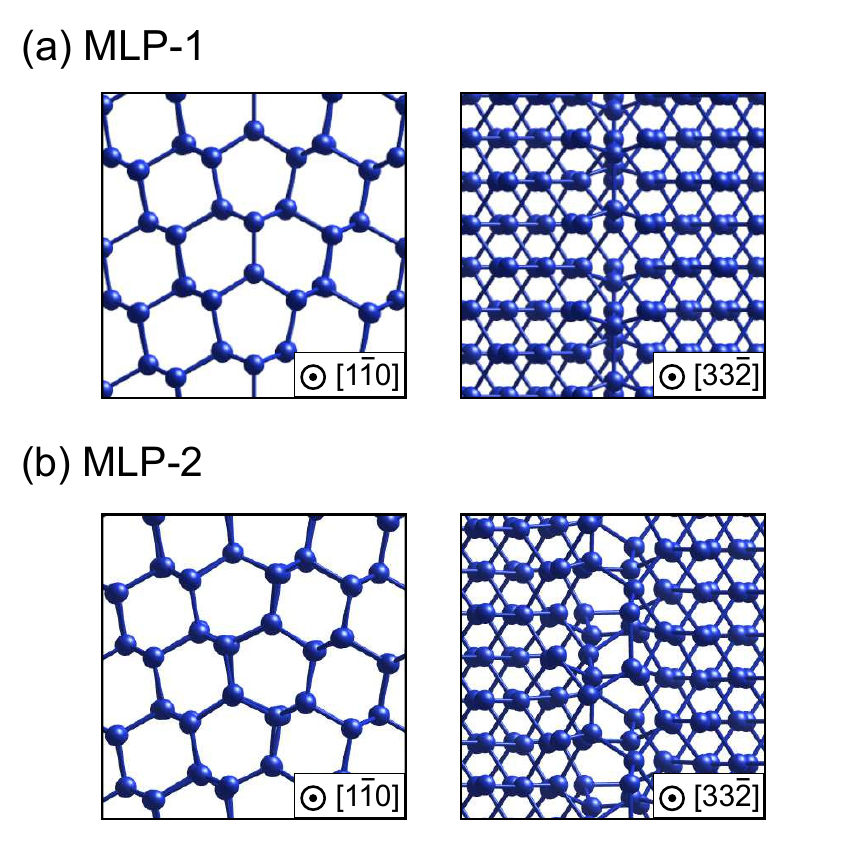}
\caption{
(a) RBT(MLP-1) and (b) RBT(MLP-2) structures for $\Sigma11$(113) GB.
}
\label{SF-AS:Fig-str-comp}
\end{figure}

We also explore the stable GB structures using the RBT approach and MLP-2.
Note that the RBT(MLP-1) and RBT(MLP-2) structures are different only for $\Sigma11 (113)$ GB.
The RBT(MLP-1) and RBT(MLP-2) structures for $\Sigma11 (113)$ GB are shown in Fig. \ref{SF-AS:Fig-str-comp}.
The GB energy of the RBT(MLP-2) structure predicted by MLP-2 is 15 mJ/m$^2$ lower than that of the RBT(MLP-1) structure.
Thus, MLP-1 has a high predictive power for most of the GB structures, and even a slight modification of MLP-1 improves the predictive power for the GB energy.

The phonon properties and lattice thermal conductivity predicted using MLP-2 and the other Pareto optimal MLPs are shown in Fig. \ref{SF-AS:Fig-convergence}.
As well as MLP-1, MLP-2 reproduces the experimental \cite{Glassbrenner1964} and DFT-calculated lattice thermal conductivities for monocrystalline silicon more quantitatively than the empirical potentials, which indicates that MLP-2 exhibits a high predictive power for the phonon properties and lattice thermal conductivity.
Therefore, MLPs should predict phonon properties and lattice thermal conductivity at GBs accurately.

\begin{figure*}[tbp]
\includegraphics[clip,width=0.9\linewidth]{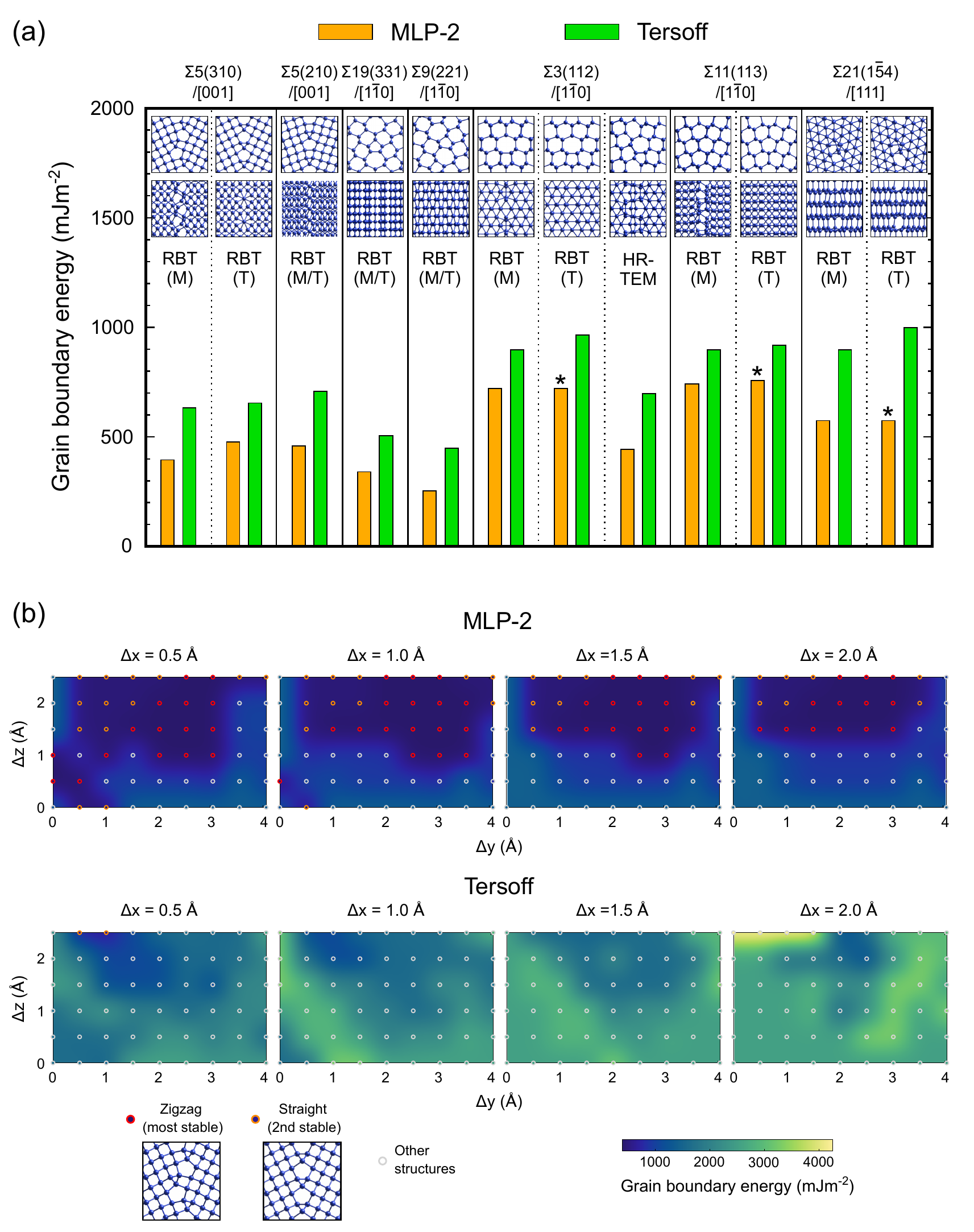}
\caption{
(a) RBT(MLP-2) and RBT(Tersoff) structures along with their GB energies predicted using MLP-2 and the Tersoff potential.
RBT(M) and RBT(T) indicate RBT(MLP-2) and RBT(Tersoff) structures (the most stable structures predicted by a rigid-body translation approach using MLP-2 and the Tersoff potential), respectively.
The asterisks indicate that the RBT(MLP-2) structures are obtained from the local geometry optimization starting from the RBT(Tersoff) structures.
The result of the HRTEM structure for $\Sigma3(112)$ \cite{Sakaguchi2007} is also shown.
(b) GB energy surfaces for $\Sigma5(310)$ GB as a function of RBTs obtained from MLP-2 and the Tersoff potential.
The RBTs deriving the zigzag RBT(MLP-2) and straight RBT(Tersoff) structures are indicated by the red and orange circles, respectively.
}
\label{SF-AS:Fig-gb-energy-rbt}
\end{figure*}

Figure \ref{SF-AS:Fig-gb-energy-rbt}(a) shows the RBT(MLP-2) and RBT(Tersoff) structures along with their GB energies predicted using MLP-2 and the Tersoff potential.
For $\Sigma5(210)$, $\Sigma19(331)$, and $\Sigma9(221)$ GBs, both MLP-2 and the Tersoff potential derive the same GB structures.
On the other hand, the RBT(MLP-2) and RBT(Tersoff) structures are different for $\Sigma5(310)$, $\Sigma3(112)$, $\Sigma11(113)$, and $\Sigma21(1\bar54)$ GBs.
The RBT(MLP-2) structures exhibit lower GB energies than the RBT(Tersoff) structures, even when predicting their GB energies using the Tersoff potential, which means that the Tersoff potential often fails to find stable GB structures.
Also, lattice dynamics calculations using the Tersoff potential show no imaginary phonon frequencies for the RBT(Tersoff) structures. 
In contrast, imaginary phonon frequencies are found in lattice dynamics calculations for the RBT(Tersoff) structures of $\Sigma 3(112)$ and $\Sigma 11(113)$ GBs using MLP-2, which implies that more stable GB structures exist.
Moreover, the RBT(MLP-2) structure of $\Sigma5(310)$ GB is of the zigzag type, whereas the RBT(Tersoff) structure is of the straight type.
The former corresponds to the GB structure observed in diamond-type germanium by high-resolution electron microscopy \cite{Bourret1988} and was found to be more stable than the straight GB structure in previous studies by DFT calculations \cite{Maiti1996,Shi2010}.
These results demonstrate that MLP-2 is more beneficial for finding GB structures accurately than the Tersoff potential, which enables us to find GB structure--GB property relationships.

\subsection{GB energy surface}

The usefulness of MLP-2 can also be found in the GB energy surface, representing the relationship between the GB energy and the RBT.
Figure \ref{SF-AS:Fig-gb-energy-rbt}(b) shows the GB energy surface of $\Sigma5(310)$ GB obtained using MLP-2 and the Tersoff potential.
As shown in Fig. \ref{SF-AS:Fig-gb-energy-rbt}(b), MLP-2 derives the zigzag RBT(MLP-2) structure for many RBTs, even a large RBT in the $x$-direction.
On the other hand, the straight RBT(Tersoff) structure is found by the RBT approach using the Tersoff potential only for two types of RBT ($\Delta x = 0.5$ \AA, $\Delta y = 0.5$ or $1.0$ \AA\ and $\Delta z = 2.5$ \AA).
Moreover, the GB energy surface obtained using the Tersoff potential has a broad region showing high GB energy for nonoptimal GB structures with large separations in the $x$-direction, whereas the GB energy surface obtained using MLP-2 is smoother than that obtained using the Tersoff potential and provides reasonable GB energies of nonoptimal GB structures.
This considerable stability of finding the GB structure originates from the fact that MLP-2 was developed from a wide range of crystal structures with large displacements and distortions (see Fig. \ref{SF-AS:Fig-selected-mlp}).
The current results also indicate that the MLP should be advantageous in modeling polycrystalline silicon where various GB structures are needed to be optimized simultaneously.

\begin{figure*}[htbp]
\includegraphics[clip,width=\linewidth]{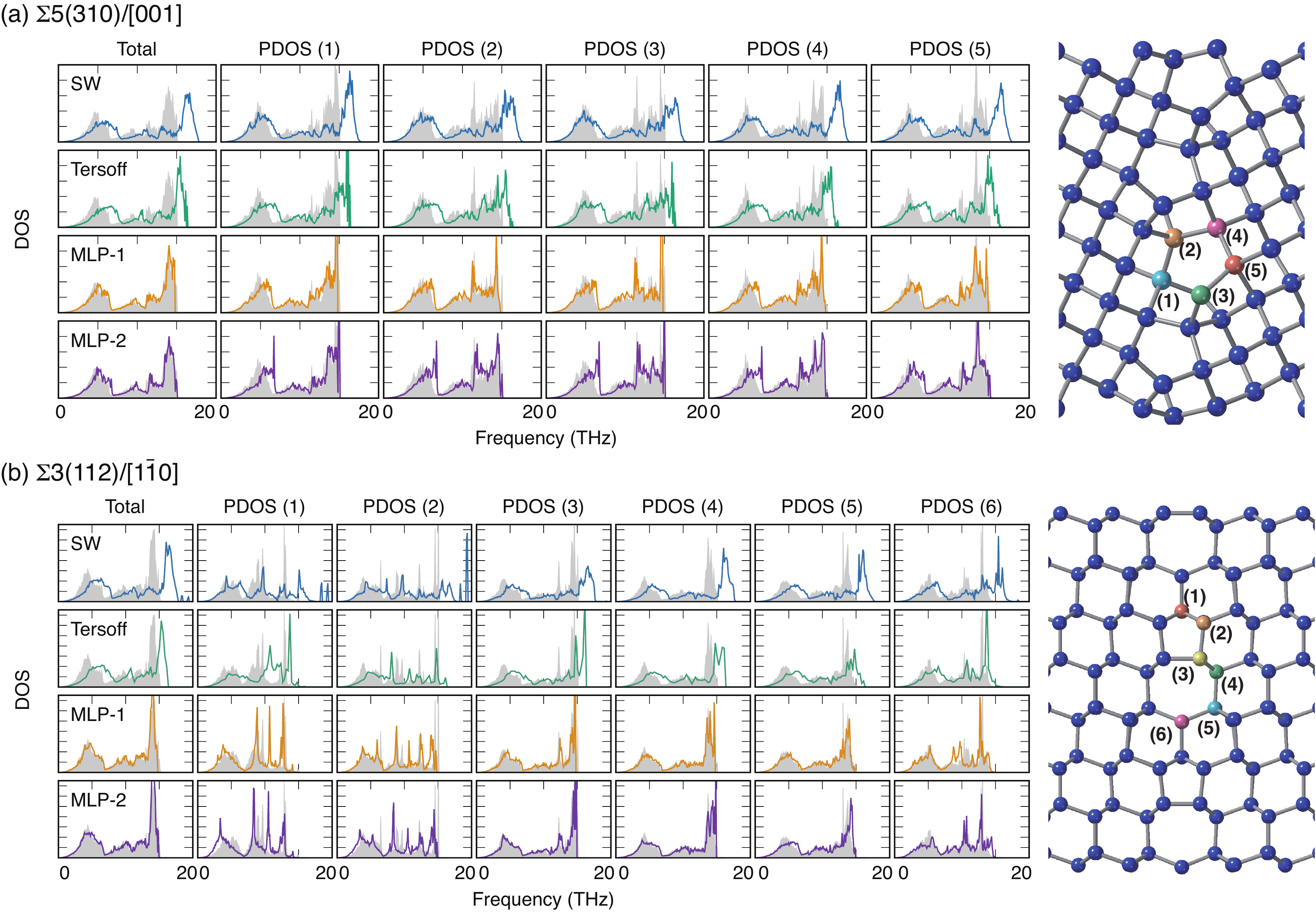}
\caption{
Projected phonon DOS for sites on the GB plane of (a) $\Sigma5(310)$ and (b) $\Sigma3(112)$ GB structures, predicted using the SW potential, Tersoff potential, MLP-1, and MLP-2.
The total phonon DOS for the GB structures is also shown.
The shaded region indicates the phonon DOS computed by DFT calculations.
}
\label{SF-AS:Fig-gb-phonon-pdos}
\end{figure*}

\subsection{Phonon properties at GBs}

We also compute the phonon density of states (DOS) for the RBT(MLP-2) structures of $\Sigma5(310)$ and $\Sigma3(112)$ GBs by DFT calculations to demonstrate the predictive power of MLP-1 and MLP-2 for phonon-related properties at GBs.
Figures \ref{SF-AS:Fig-gb-phonon-pdos}(a) and (b) show the projected phonon DOS for five sites on the GB plane of the RBT(MLP-2) structure of $\Sigma5(310)$ GB and those for six sites on the GB plane of the RBT(MLP-2) structure of $\Sigma3(112)$ GB, respectively.
As can be seen in Figs. \ref{SF-AS:Fig-gb-phonon-pdos}(a) and (b), the SW and Tersoff potentials tend to overestimate the phonon frequencies of the GB structures.
On the other hand, the total and projected phonon DOS computed using MLP-1 and MLP-2 are very close to those computed using the DFT calculation.
MLP-1 and MLP-2 predict phonon properties at GBs accurately, although MLP-1 was developed without using GB structures.

As shown thus far, MLP-1 shows high predictive power for the GB structures and properties other than the absolute values of GB energy.
Therefore, MLP-1 should predict the properties of the GB structures almost by interpolation.
Then, MLP-2 can be regarded as a slightly modified version of MLP-1, such that MLP-2 can be used in the prediction of GB energy.
For the following thermal conduction analyses, we employ MLP-2 providing high predictive power for GB energy.

\section{Thermal conduction at grain boundaries}
\subsection{Perturbed molecular dynamics simulation}

\newcommand{\vi}{{\bf v}_i}
\newcommand{\ri}{{\bf r}_i}
\newcommand{\rij}{{\bf r}_{ij}}
\newcommand{\fij}{{\bf f}_{ij}}
\newcommand{\Fexts}{F_{\rm ext}}
\newcommand{\Phiij}{{\it \Phi}_{ij}}

We calculate the lattice thermal conductivity across the GB planes using the perturbed MD method \cite{Ciccotti1979,Evans1982,Gillan1983,Yoshiya2004}.
We use an implementation of the perturbed MD for the \textsc{lammps} code developed by one of the authors\cite{Fujii2018}.
For a given computational model including GB planes elongated in a single direction ($x$-direction), the lattice thermal conductivity in the $x$-direction at the absolute temperature $T$ can be calculated by the perturbed MD simulation as
\begin{equation}
\kappa = \frac{1}{\Fexts T} \lim_{t \to \infty} {\langle J_x \rangle}_{t},
\end{equation}
where $\Fexts$ and $J_x$ denote the given magnitude of the perturbation and the heat flux in the $x$-direction, respectively.
The three-dimensional microscopic heat flux $\bf J$ in the GB model with volume $V$ is then given as \cite{Irving1950} 
\begin{equation}
{\bf J} = \frac{1}{2V} \sum_{i} \left[ \left\{ m_i \vi^{2} {\bf I} + \sum_{j} \Phiij {\bf I} \right\} \cdot \vi - \sum_{j} \left( \fij \cdot \vi \right) \rij \right],
\end{equation}
where $\Phiij$ and $\fij$ denote the interatomic potential energy between atoms $i$ and $j$ and the force exerted by atom $j$ on atom $i$, respectively.
$\bf I$ is the unit tensor of the second rank, and $m_i$ and $\vi$ are the mass and velocity of atom $i$, respectively.

First, we estimate the dependence of the thermal conductivity of monocrystalline diamond-type silicon on simulation cell size using the $4\times4\times4$, $5\times5\times5$, $6\times6\times6$ and $7\times7\times7$ expansions of its unit cell with the lattice parameter of 5.466 {\AA}.
The temperature is set to 700 K, which is higher than the Debye temperature of diamond silicon \cite{Keyes1978}.
Thus, any quantum effects on the lattice thermal conductivity should be negligible.
The monocrystalline thermal conductivity converges well at the $4\times4\times4$ supercell for MLP-2 to $\sim$38 W/mK and at the $6\times6\times6$ supercell for the Tersoff potential to $\sim$99 W/mK.

Then, we construct a computational model for each GB structure, including two GB planes separated by approximately 5 nm for the $x$-axis.
According to the finite cell size dependence of thermal conductivity, cells with sides of 22 {\AA} or longer are used for MLP-2, and those with sides of 33 {\AA} or longer are used for the Tersoff potential.
Thus, the GB models are composed of 4320 -- 4640 atoms for MLP-2 and 5568 -- 7200 atoms for the Tersoff potential.
Before performing the perturbed MD simulation, standard MD simulations with the NPT ensemble are carried out for 100 ps with a time step of 1 fs (100,000 steps) to determine the equilibrium cell dimensions of each GB model at 700 K.
Then, the atomic configuration of the GB model is equilibrated efficiently by NVT-MD simulation with a temperature scaling for 100 ps (100,000 steps) and NVT-MD simulation with the Nos\'e--Hoover thermostat for 300 ps (300,000 steps).

Using the GB models in the equilibrium state, we perform the perturbed MD simulations for 1100 ps (1,100,000 steps).
The average heat flux required for calculating the lattice thermal conductivity is then estimated from the trajectory during the last 1000 ps after reaching the steady state of the perturbed GB model.
We apply five different magnitudes of the perturbation for each of the GB models, and the average and standard deviation of thermal conductivities for the five simulations are calculated.
Note that we have confirmed that the current magnitudes of perturbation can be within the linear response regime.
Further details on the formulation of the perturbed MD method are available elsewhere \cite{Yoshiya2004,Fujii2014}.
When using the Tersoff potential, which is much less computationally demanding, we set the simulation time to 1000 ps for NPT-MD, 1000 ps for NVT-MD with temperature scaling, 3000 ps for NVT-MD with the Nos\'e--Hoover thermostat, and 3100 ps for the perturbed MD. 
The use of longer MD simulation times should improve the statistical accuracy of lattice thermal conductivity.

Finally, we estimate the GB thermal resistance ${R_{\rm GB}}$ for each GB model from the lattice thermal conductivity calculated as \cite{Yang2002}
\begin{equation}
\kappa_{\rm GB} = \frac{\kappa_{\rm bulk}}{1 + R_{\rm GB}\kappa_{\rm bulk}/d},
\end{equation}
where $\kappa_{\rm GB}$ and $\kappa_{\rm bulk}$ are the lattice thermal conductivity of the GB model with the separation between two GB planes $d$ and that of the bulk crystal, respectively.

\subsection{Phonon wave packet method}

We also investigate the process of phonon scattering at GBs using the phonon wave packet method \cite{Schelling2002}.
This method begins with a wave packet of phonons generated from a phonon mode in the bulk crystal.
Initially, the wave packet propagates at its group velocity.
Part of the wave packet is then transmitted through the GB plane and the remaining is reflected at the GB plane involving the mode conversion.
This method is beneficial for elucidating the behavior of each phonon at the GB plane, while phonon--phonon interactions and anharmonic effects at finite temperatures are not generally included in contrast to the perturbed MD simulation.

Here, we consider only the phonons with the wave vector normal to the GB plane because they dominate the thermal conduction across interfaces \cite{Deng2014}.
For the phonon wave packet simulations, a supercell composed of two crystals forming a GB plane should be remarkably long for the direction perpendicular to the GB plane ($x$-direction), which enables us to analyze the reflected and transmitted phonons that propagate at the group velocities different from that of the initial wave packet.
When using MLP-2 and the Tersoff potential, we set the length of one crystal (the number of atoms) in a supercell to approximately 5000 {\AA} (22966 -- 35452 atoms) and 15,000 {\AA} (69846 -- 108220 atoms), respectively.
These structures are relaxed until the residual forces are less than at least $3.0 \times 10^{-7}$ eV/\AA.
The current wave packet simulation can be completed only for a limited number of given phonon modes because some reflected and transmitted phonons reach the edges of crystals before the scattering process at the GB finishes.
Moreover, we omit wave packet simulations using MLP-2 on the basis of phonon modes with very low group velocities because the required MD steps of more than 100,000 are very large.

In the wave packet simulation, atomic displacements associated with a phonon wave packet are expressed as \cite{Schelling2002,Schelling2004}
\begin{widetext}
\begin{equation}
{\bf u}_{il} = A{\bf \epsilon}_{i\lambda{\bf k}}{\rm exp}[ik_x(x_{il}-x_0)]{\rm exp}[-\eta^2(x_{il}-x_0)^2]{\rm exp}(-i\omega_{\lambda{\bf k}}t),
\label{SF-AS:Eq-pwp}
\end{equation}
\end{widetext}
where ${\bf u}_{il}$ is the displacement vector of atom $i$ in the primitive cell $l$ and $A$ is the amplitude of the wave packet. 
$\omega_{\lambda{\bf k}}$ and ${\bf \epsilon}_{i\lambda{\bf k}}$ denote the eigenvalue and eigenvector of atom $i$ for the phonon mode identified with the wave vector ${\bf k}$ and branch $\lambda$, respectively.
$k_x$, $x_{il}$, and $x_0$ are the $x$ component of ${\bf k}$, that of the position of atom $i$ in primitive cell $l$, and that of the center of the wave packet, respectively. 
$\eta$ is the factor that controls the spatial width of the wave packet and $t$ is the time.
Initial atomic displacements for a wave packet are generated with $t = 0$, and initial atomic velocities are obtained using the time derivative of Eq. (\ref{SF-AS:Eq-pwp}).
In generating the initial atomic displacements and velocities, we set $A = 10^{-3} / {\sqrt{m_i}} $ {\AA} ($m_i = 28.0855$ for silicon) and $\eta = 0.006$ \AA$^{-1}$.
The eigenvectors and eigenvalues of phonons in the bulk crystal are obtained using \textsc{phonopy} \cite{phonopy2}.
The transmission coefficient $\alpha_{\lambda{\bf k}}$ of a wave packet with wave vector ${\bf k}$ and branch $\lambda$ is calculated from the partition of kinetic energy into two neighboring crystals after the scattering process was completed \cite{Ju2013}.

From the result of phonon wave packet simulations, the GB thermal resistance ${R_{\rm GB}}$ and GB thermal conductance ${\sigma_{\rm GB}}$ can be estimated as \cite{Young1989,Zheng2014}
\begin{equation}
\frac{1}{R_{\rm GB}} = {\sigma_{\rm GB}} = \frac{1}{V}\sum_{\lambda,{\bf k}}^{+}C_{\lambda{\bf k}}v_{\lambda{\bf k}x}\alpha_{\lambda{\bf k}},
\label{SF-AS:Eq-pwp-res}
\end{equation}
where $V$ is the system volume, and $C_{\lambda{\bf k}}$ and $v_{\lambda{\bf k}x}$ are respectively the mode-dependent heat capacity and the $x$ component of group velocity with the wave vector ${\bf k}$ and branch $\lambda$.
Only phonons traveling toward the GB (i.e., positive $v_{\lambda{\bf k}x}$) are summed over in this equation.
This summation should be carried out for all phonon modes in the first Brillouin zone.
However, because we perform wave packet simulations for only phonons with the wave vector normal to the GB plane, it is difficult to reliably estimate the absolute GB thermal resistance.
Thus, we obtain the ratio of the GB thermal resistance of one GB atomic structure to that of another structure with the same GB plane [RBT(MLP-2) and RBT (Tersoff) for $\Sigma5(310)$; RBT(MLP-2) and HRTEM for $\Sigma3(112)$].
The GB thermal resistances of these two structures should be comparable because they are calculated from the same phonon wave packets.

\subsection{Results and discussion}

\begin{figure}[tbp]
\includegraphics[clip,width=\linewidth]{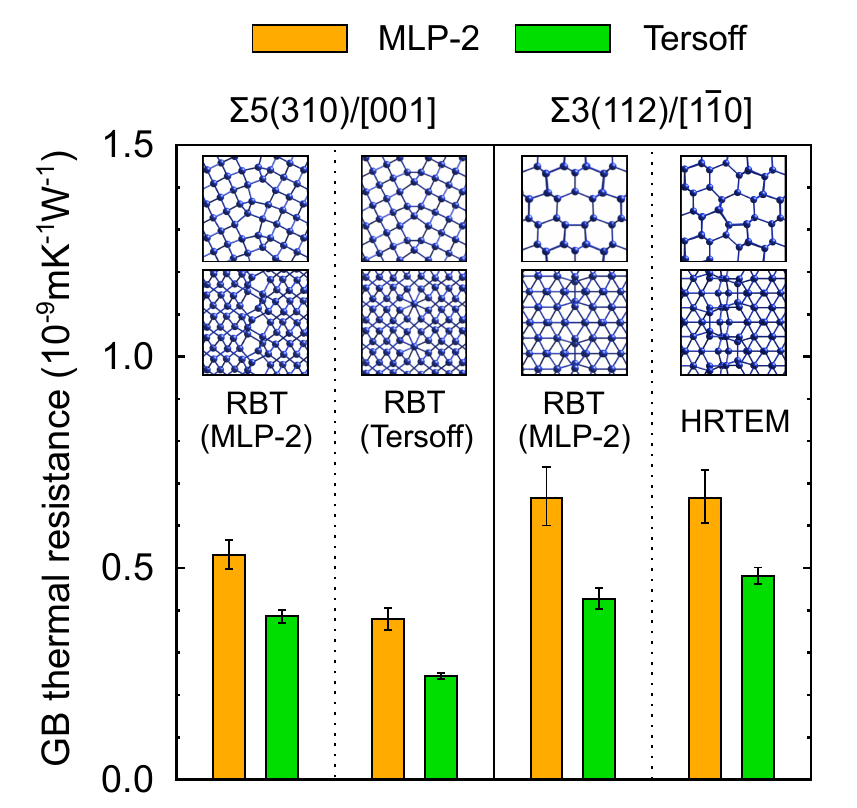}
\caption{
GB thermal resistances for $\Sigma5(310)$ and $\Sigma3(112)$ GBs calculated by perturbed MD simulations using MLP-2 and Tersoff potential.
}
\label{SF-AS:Fig-resistance}
\end{figure}

Figure \ref{SF-AS:Fig-resistance} shows the GB thermal resistances of four GB structures calculated by the perturbed MD simulation, namely, the RBT(MLP-2) and RBT(Tersoff) structures for $\Sigma 5(310)$ GB and the RBT(MLP-2) and HRTEM structures \cite{Sakaguchi2007} for $\Sigma 3(112)$ GB.
They are predicted using both MLP-2 and the Tersoff potential.
As found in Fig. \ref{SF-AS:Fig-resistance}, the GB thermal resistance depends on the GB plane and GB microscopic structure.
Firstly, the GB thermal resistances for $\Sigma 3(112)$ GB are higher than those for $\Sigma 5(310)$ GB when predicting them using either MLP-2 or the Tersoff potential.
Regarding the microscopic GB structure dependence, the GB resistances of the RBT(MLP-2) and HRTEM structures for $\Sigma 3(112)$ GB are similar.
On the other hand, the GB resistance of the RBT(Tersoff) structure for $\Sigma 5(310)$ is lower than that of the RBT(MLP-2) structure, possibly because of the symmetric atomic configuration of the RBT(Tersoff) structure.
The GB structure dependence of the GB resistance predicted using the Tersoff potential and that predicted using MLP-2 are similar, although the Tersoff potential produces lower GB thermal resistances than MLP-2.
Moreover, the GB thermal resistance shows no correlation with the GB energy in the current results.
The RBT(Tersoff) structure for $\Sigma 5(310)$ GB and the HRTEM structure for $\Sigma 3(112)$ GB have similar GB energies, whereas their GB thermal resistances differ by a factor of two (see also Fig. \ref{SF-AS:Fig-gb-energy-rbt}).

\begin{figure}[tbp]
\includegraphics[clip,width=\linewidth]{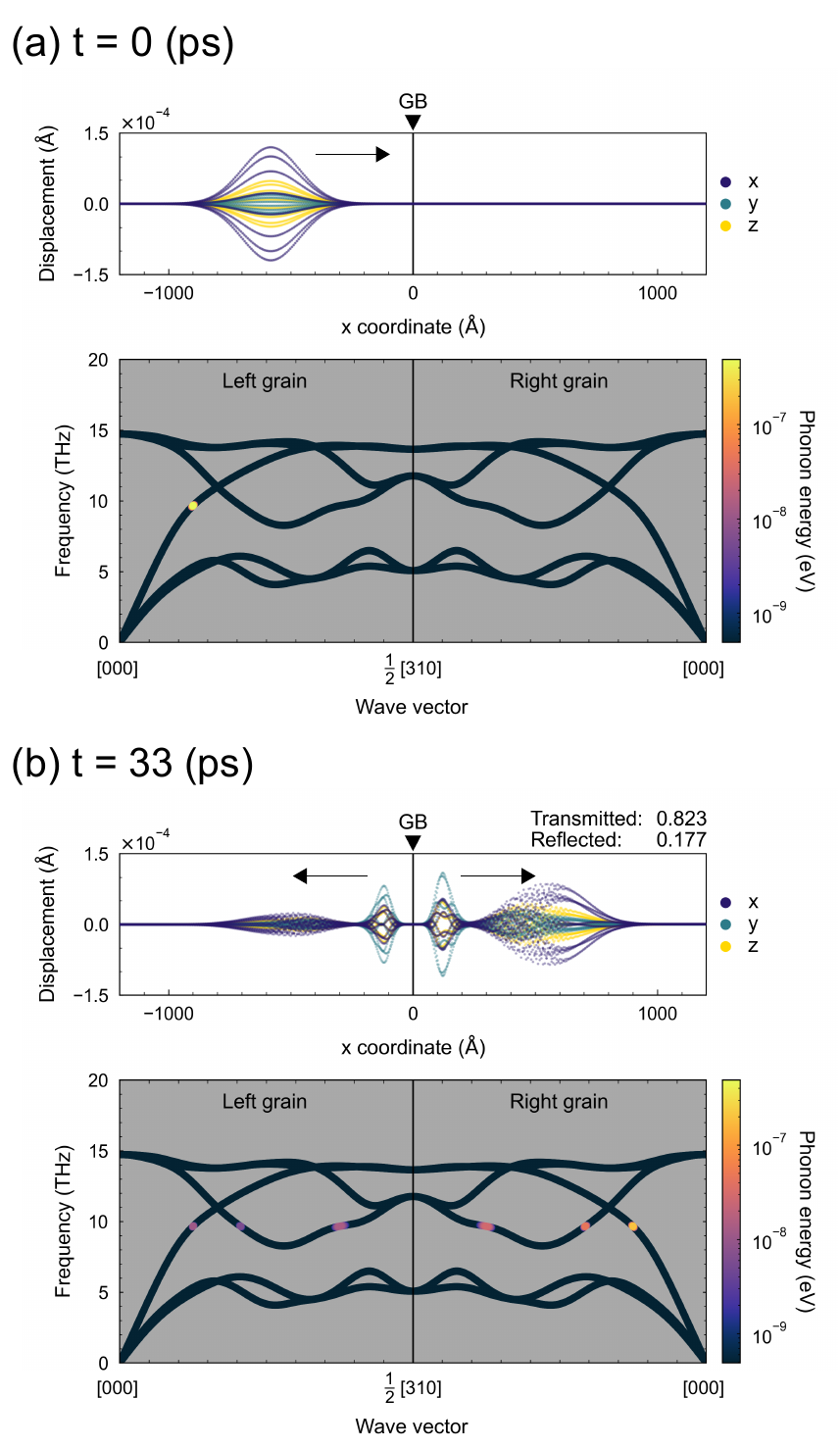}
\caption{ 
Example of phonon wave packet simulation using MLP-2 for the RBT(MLP-2) structure of $\Sigma 5(310)$ GB at (a) $t$ = 0 ps and (b) $t$ = 33 ps.
The upper panels in (a) and (b) show the atomic displacements associated with a wave packet from phonon mode ${\bf k} = 1/8[310]$.
The lower panels show the decomposition of the atomic displacements into phonon modes and the phonon energies estimated from the corresponding phonon amplitude.
Note that, although only phonons with the wave vectors parallel to the GB plane are plotted in the lower panels, phonons with the other wave vectors in the entire Brillouin zone are also found in the decomposition.
}
\label{SF-AS:Fig-pwp}
\end{figure}

Figure \ref{SF-AS:Fig-pwp} shows an example of the phonon wave packet simulation performed using MLP-2 and the RBT(MLP-2) structure of $\Sigma 5(310)$ GB.
A wave packet associated with one specific phonon mode is generated in the left grain as shown in Fig. \ref{SF-AS:Fig-pwp}(a). 
A wave packet is represented by a linear combination of phonon modes identified with plane waves centered at one specific phonon mode.
Then, it propagates to the GB plane at the group velocity that corresponds to the gradient of the phonon dispersion at the wave vector of the phonon mode.
Finally, the wave packet is transmitted or reflected at the GB, involving the conversion to phonon modes with different wave vectors as shown in Fig. \ref{SF-AS:Fig-pwp}(b).

\begin{figure*}[tbp]
\includegraphics[clip,width=\linewidth]{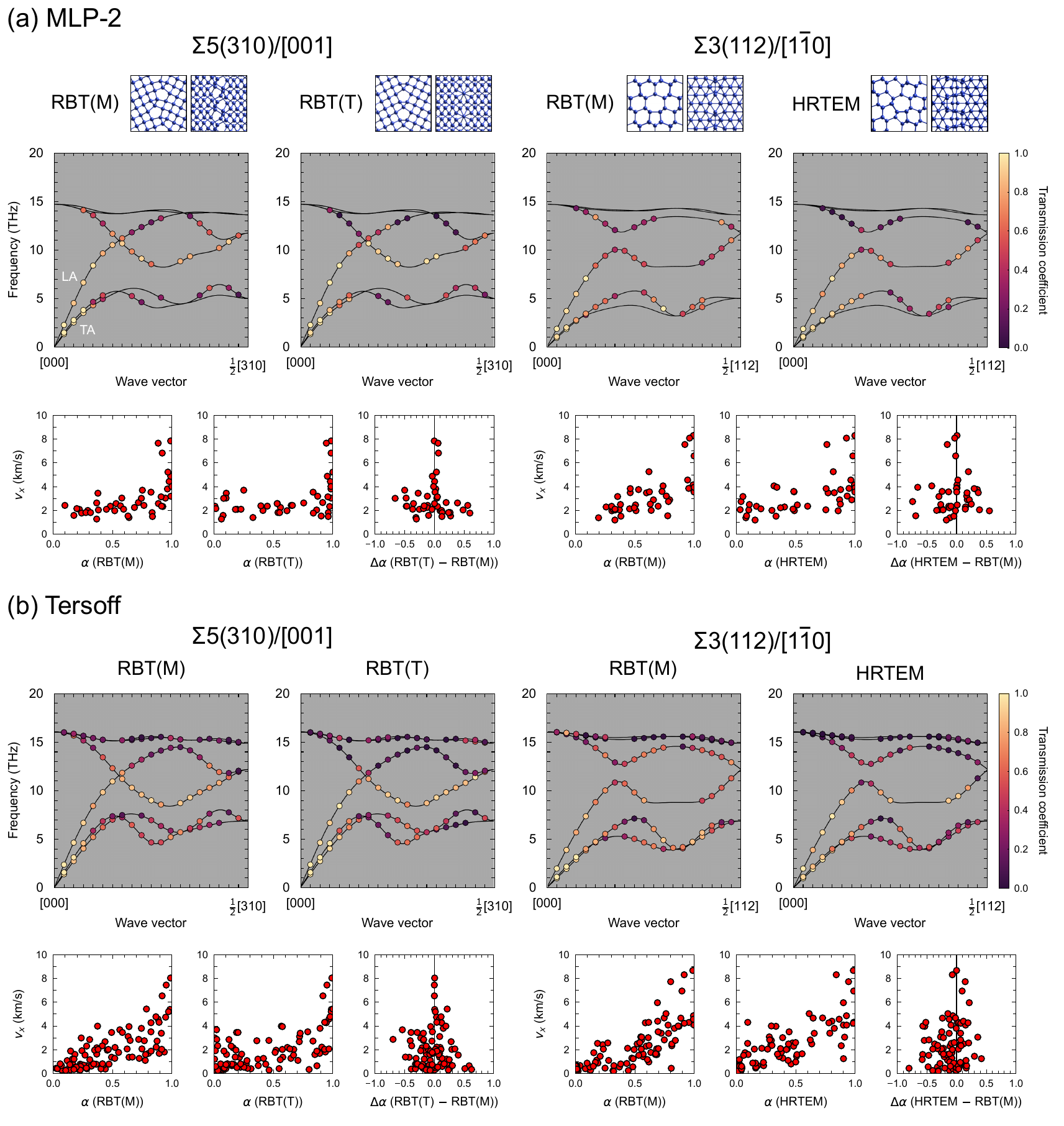}
\caption{
Transmission coefficient of a wave packet centered at a wave vector perpendicular to the GB plane obtained using (a) MLP-2 and (b) the Tersoff potential.
The results of the RBT(MLP-2) and RBT(Tersoff) structures for $\Sigma 5(310)$ GBs and the RBT(MLP-2) and HRTEM structures \cite{Sakaguchi2007} for $\Sigma 3(112)$ GBs are shown.
The transmission coefficients are shown along the phonon dispersions in the upper panels, whereas their relationships with group velocity is shown in the lower panels.
For the phonons with $v_{\lambda{\bf k}x} < 0$ (i.e., the phonon does not propagate toward the GB), the results of $\alpha_{\lambda-{\bf k}}$ are depicted instead of those of $\alpha_{\lambda{\bf k}}$.
RBT(M) and RBT(T) indicate the most stable structures predicted by a rigid-body translation approach using MLP-2 and the Tersoff potential, respectively.
}
\label{SF-AS:Fig-transmission}
\end{figure*}

This procedure is systematically applied to the other phonon modes with the wave vectors perpendicular to the GB plane.
Figure \ref{SF-AS:Fig-transmission}(a) shows the dependence of the transmission coefficient on the phonon mode predicted using MLP-2.
There is a trend that the wave packets with high group velocities (i.e., large momentum perpendicular to the GB plane) exhibit large transmission coefficients.
In particular, longitudinal-acoustic (LA) and transverse-acoustic (TA) modes, which mainly contribute to the thermal conductivity in bulk silicon \cite{Broido2007,Esfarjani2011}, are mostly transmitted for the four GB atomic structures. 
Therefore, they should dominate the GB thermal conductance.

\begin{table}[tbp]
\begin{ruledtabular}
\caption{
Ratio of GB thermal resistance of one GB atomic structure to that of another structure with the same plane calculated by phonon wave packet (PWP) and perturbed MD (PMD) simulations. The ratios obtained using MLP-2 and the Tersoff potential are shown.
}
\label{SF-AS:Table-kappa}
\begin{tabular}{lcccc}
& \multicolumn{2}{c}{$\Sigma5(310)/[001]$} & \multicolumn{2}{c}{$\Sigma3(112)/[1\bar10]$} \\
& \multicolumn{2}{c}{\scriptsize{$R$(RBT(T))/$R$(RBT(M))}} & \multicolumn{2}{c}{\scriptsize{$R$(HRTEM)/$R$(RBT(M))}} \\
\cline{2-3}\cline{4-5}
& MLP-2 & Tersoff & MLP-2 & Tersoff \\
\hline
PWP & 1.04 & 1.05 & 1.05 & 1.14 \\
PMD & 0.71 & 0.63 & 1.00 & 1.06 \\
\end{tabular}
\end{ruledtabular}

\end{table}

The lower panels of Fig. \ref{SF-AS:Fig-transmission}(a) show the differences in transmission coefficients of identical phonon modes between two GB atomic structures of the same GB plane ($\Delta\alpha$).
In contrast to LA and TA modes in the neighborhood of the $\rm{\Gamma}$ point, the transmission coefficients of phonons with lower group velocities are highly dependent on the GB atomic structures.
However, distributions of the transmission coefficients against the group velocities are similar between the RBT(MLP-2) and RBT(Tersoff) structures for $\Sigma 5(310)$ GB and between the RBT(MLP-2) and HRTEM structures for $\Sigma 3(112)$ GB.
As a result, two different GB atomic structures with the same GB plane exhibit comparable GB thermal resistances for both $\Sigma 5(310)$ and $\Sigma 3(112)$ GBs in the phonon wave packet simulations (the ratio is $\sim$1), as summarized in Table \ref{SF-AS:Table-kappa}.
For $\Sigma 5(310)$ GBs, the ratios obtained from phonon wave packet simulations are inconsistent with those obtained from perturbed MD simulations, where the RBT(Tersoff) structure exhibit a 0.7 times lower thermal resistance than the RBT(MLP-2) structure (Table \ref{SF-AS:Table-kappa}).
Considering that MD simulations include any anharmonic effects of atomic vibrations at a finite temperature, whereas phonon wave packet simulations do not include phonon--phonon interactions, the difference in GB thermal resistances originates from anharmonic vibrations in the vicinity of GBs.

Figure \ref{SF-AS:Fig-transmission}(b) shows the transmission coefficients calculated using the Tersoff potential.
The phonon mode dependence predicted using the Tersoff potential is qualitatively consistent with that predicted using MLP-2, except that the HRTEM structure of $\Sigma 3(112)$ GB has a slightly higher thermal resistance than the RBT(MLP-2) structure (Table \ref{SF-AS:Table-kappa}).
The similar results obtained using MLP-2 and the Tersoff potential are probably because of the fact that the phonon transmission and reflection at the GB planes are chiefly determined by the geometric structure of the GB.

\section{Conclusion}
We have investigated the GB structures and GB phonon properties in silicon using MLPs.
The current Pareto optimal MLPs taken from our repository are composed of polynomial invariants of the O(3) group representing the neighboring atomic density.
This study shows that the MLPs exhibit high predictive power for the GB excessive energy and phonon properties at GBs.
Furthermore, the MLPs provide reasonable GB energies of nonoptimal GB structures, which enables us to perform a robust and efficient search for optimal GB structures.

We have also examined the lattice thermal conduction behavior using a combination of MLPs, perturbed MD, and phonon wave packet method with simulation cells containing an enormously large number of atoms and with many MD simulation steps.
The GB thermal resistances estimated by the perturbed MD and wave packet simulations show an important role of anharmonic vibrations at GBs in determining interfacial thermal conduction, in addition to phonon transmissions and reflections at GBs.

The comparison between the MLP and the Tersoff potential has revealed two main advantages of the MLP for GB thermal conduction analyses: (1) the MLP derives the optimal GB structure even when the Tersoff potential fails.
As the GB thermal resistance depends on the microscopic GB structure, this feature is necessary for elucidating a reliable orientation--structure--property relationship.
(2) The MLP well reproduces the experimental/DFT monocrystalline thermal conductivity and phonon DOS for the bulk and GB structures.
Thus, the current MLP exhibits reliable thermal conduction behaviors at GBs.

\begin{acknowledgments}
SF was supported by a Grant-in-Aid for Early-Career Scientists (Grant Number JP20K15034) and Grants-in-Aid for Scientific Research on Innovative Areas (Grant Numbers JP20H05195 and JP19H05786) from the Japan Society for Promotion of Science (JSPS).
AS acknowledges a Grant-in-Aid for Scientific Research (B) (Grant Number 19H02419) and a Grant-in-Aid for Scientific Research on Innovative Areas (Grant Number 19H05787) from JSPS.
\end{acknowledgments}

\appendix

\section{Current MLPs}
\label{SF-AS:Appendix-MLP}
\subsection{Potential energy model}
The current MLPs (MLP-1 and MLP-2) and the other MLPs in the repository \cite{MachineLearningPotentialRepositoryArxiv,MachineLearningPotentialRepository} are composed of linearly independent polynomial invariants of the O(3) group. 
The polynomial invariants are derived from order parameters representing the neighboring atomic density in terms of radial functions and spherical harmonics \cite{PhysRevB.99.214108,PhysRevB.102.174104}.
A $p$th-order polynomial invariant for a given radial number $n$ and a given set of angular numbers $\set{l_1,l_2,\cdots,l_p}$ is defined by a linear combination of products of $p$ order parameters, expressed as
\begin{equation}
\label{Eqn-invariant-form}
d_{nl_1l_2\cdots l_p, (\sigma)}^{(i)} = \sum_{m_1, \cdots, m_p} C^{l_1l_2\cdots l_p, (\sigma)}_{m_1m_2\cdots m_p} a_{nl_1m_1}^{(i)} a_{nl_2m_2}^{(i)} \cdots a_{nl_pm_p}^{(i)},
\end{equation}
where $a_{nlm}^{(i)}$ denotes the order parameter of the component $nlm$ representing the neighboring atomic density of atom $i$.
Note that multiple invariants for a set of $\set{l_1,l_2,\cdots,l_p}$ are distinguished by index $\sigma$ if necessary.

The current MLPs consist of polynomial invariants up to the fourth order expressed as
\begin{eqnarray}
D^{(i)} &=& D_{\rm pair}^{(i)} \cup D_2^{(i)} \cup D_3^{(i)} \cup D_4^{(i)}, \nonumber \\
D_{\rm pair}^{(i)} &=& \set{d_{n0}^{(i)}}, \nonumber \\
D_2^{(i)} &=& \set{d_{nll}^{(i)}}, \\
D_3^{(i)} &=& \set{d_{nl_1l_2l_3}^{(i)}}, \nonumber \\
D_4^{(i)} &=& \set{d_{nl_1l_2l_3l_4,(\sigma)}^{(i)}}, \nonumber
\end{eqnarray}
where $D_{\rm pair}$ and $D_p$ denote a set of pairwise invariants and a set of $p$th-order polynomial invariants, respectively.
The current MLPs adopt a finite basis set of Gaussian-type radial functions combined with a cosine-based cutoff function $f_c$ that ensures the smooth decay of the radial function and is given by
\begin{equation}
f_{n}(r)=\exp\left[-\beta_n(r-r_n)^{2}\right] f_c(r),
\end{equation}
where $\beta_n$ and $r_n$ denote the given parameters.
The order parameters for atom $i$ are then approximately calculated from its neighboring atomic density regardless of the orthonormality of the radial functions \cite{PhysRevB.99.214108} as
\begin{equation}
a_{nlm}^{(i)} = \sum_{j \in \rm {neighbor}} f_n(r_{ij}) Y_{lm}^* (\theta_{ij}, \phi_{ij}),
\label{EquationOrderParameters}
\end{equation}
where $(r_{ij}, \theta_{ij}, \phi_{ij})$ denotes the spherical coordinates of neighboring atom $j$ centered at the position of atom $i$.

Finally, in the current MLPs, the atomic energy is formulated with respect to the polynomial invariants as
\begin{equation}
E^{(i)} = F_1 \left( D^{(i)} \right)
+ F_2 \left( D_{\rm pair}^{(i)} \cup D_2^{(i)} \right),
\end{equation}
where polynomial functions with regression coefficients $\set{w}$ are given by 
\begin{eqnarray}
F_1 \left(D\right) &=& \sum_i w_{i'} d_{i'} \nonumber \\
F_2 \left(D\right) &=& \sum_{\{i',j'\}} w_{i'j'} d_{i'} d_{j'}.
\end{eqnarray}
Further details on the potential energy model using the polynomial invariants can be found in the literature \cite{PhysRevB.99.214108,PhysRevB.102.174104}.

\subsection{Predictive power for properties}

\begin{table}[tbp]
\begin{ruledtabular}
\caption{
Model parameters of MLP-1 and MLP-2.
}
\label{SF-AS:Table-model-parameter}
\begin{tabular}{lc}
number of coefficients & 4590 \\
cutoff radius (\AA) & 6.0  \\
number of radial functions & 15 \\
$\set{l_{\rm max}^{(2)}, l_{\rm max}^{(3)},l_{\rm max}^{(4)}}$ & [4,4,2] \\
RMS error (MLP-1, energy, meV/atom)  & 7.58  \\
RMS error (MLP-1, force, eV/\AA)  & 0.088   \\
RMS error (MLP-2, energy, meV/atom)  & 7.63  \\
RMS error (MLP-2, force, eV/\AA)  & 0.089   \\
\end{tabular}
\end{ruledtabular}
\end{table}

\begin{table}[tbp]
\begin{ruledtabular}
\caption{
Predicted values of properties for the diamond-type structure.
}
\label{SF-AS:Table-property}
\begin{tabular}{lcccc}
Property & SW & Tersoff & MLP-1 & DFT \\
\hline
cohesive energy (eV/atom) & 4.337 & 4.630 & 4.550 & 4.552 \\
equilibrium volume (\AA$^3$/atom) & 20.022 & 20.035 & 20.394 & 20.457  \\
elastic constants & & & & \\
$C_{11}$ (GPa) & 151.4 & 142.5 & 134.2 & 153.1  \\
$C_{12}$ (GPa) & 76.4 & 75.3 & 62.9 & 56.6  \\
$C_{44}$ (GPa) & 56.4 & 69.0 & 66.0 & 74.1  \\
\end{tabular}
\end{ruledtabular}
\end{table}

\begin{figure*}[tbp]
\includegraphics[clip,width=\linewidth]{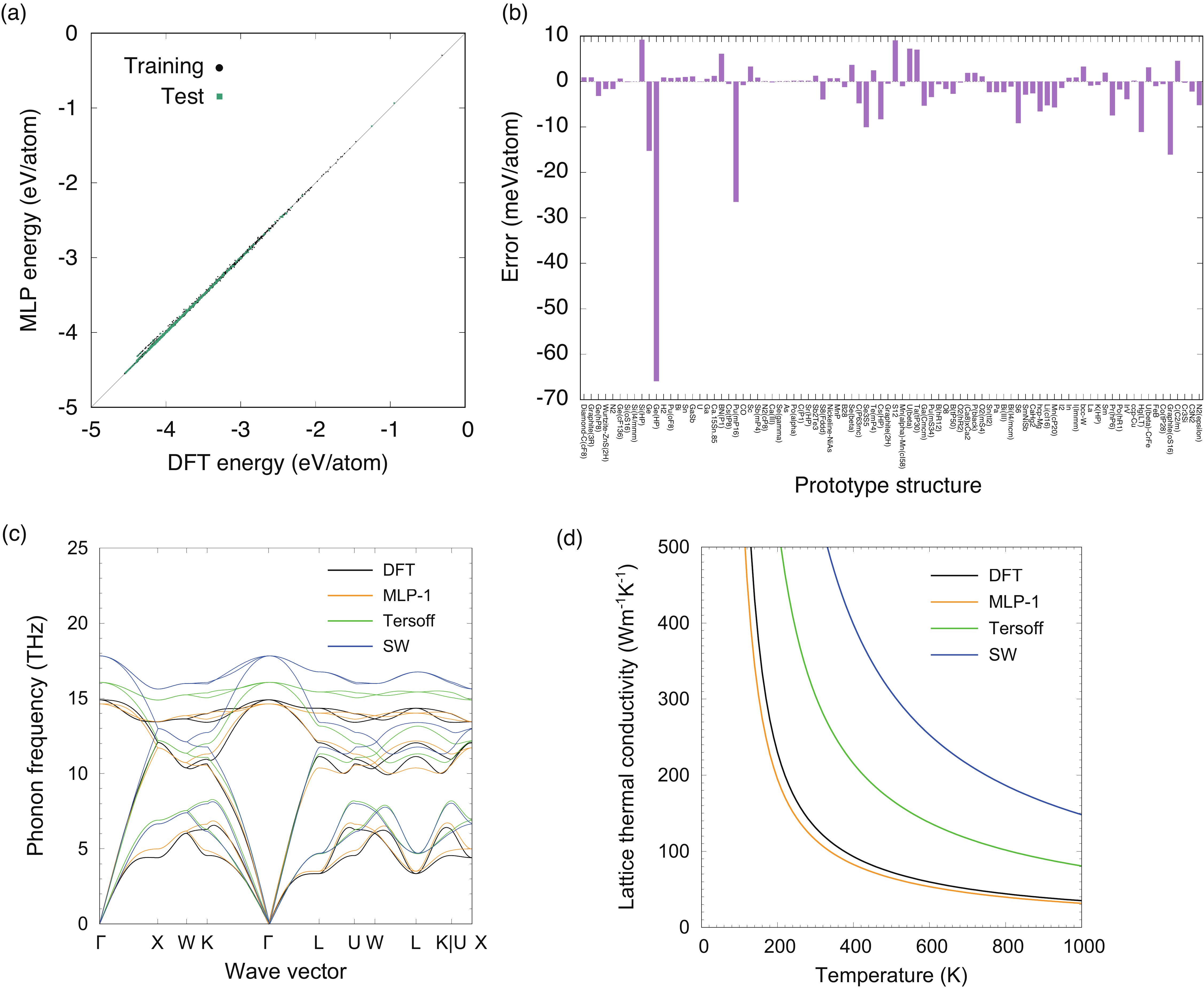}
\caption{
Properties predicted using MLP-1.
(a) Cohesive energies predicted using MLP-1 and by DFT calculations. 
(b) Structure dependence of the difference between the energy predicted by DFT calculations and that predicted using MLP-1.
For each prototype structure, the energy difference of its equilibrium structure is shown.
(c) Phonon dispersion curves predicted using MLP-1, Tersoff potential, and SW potential, and by DFT calculations.
(d) Temperature dependences of lattice thermal conductivity predicted using MLP-1, Tersoff potential, and SW potential, and by DFT calculations.
}
\label{SF-AS:Fig-selected-mlp}
\end{figure*}

Table \ref{SF-AS:Table-model-parameter} shows the model parameters used for developing MLP-1 and MLP-2.
Table \ref{SF-AS:Table-property} shows the cohesive energy, equilibrium volume, and elastic constants for the diamond-type structure predicted using the SW potential, Tersoff potential, and MLP-1, and by DFT calculations.
Except for cohesive energy, all of the potentials can predict the properties accurately.
Figure \ref{SF-AS:Fig-selected-mlp}(a) shows the distribution of the cohesive energies predicted using the current MLP and by the DFT calculations for structures in the training and test datasets.
As can be seen in Fig. \ref{SF-AS:Fig-selected-mlp}(a), the current MLP predicts the cohesive energy for most of the structures in the current datasets accurately, although it shows large systematic errors for some structures.
These systematic errors are also observed in Fig. \ref{SF-AS:Fig-selected-mlp}(b). 
Although the prediction error of the cohesive energy of some prototype structures is significant, the current MLP shows a small prediction error for a wide range of prototype structures.
Figures \ref{SF-AS:Fig-selected-mlp}(c) and (d) show the phonon dispersion curves and the temperature dependence of the lattice thermal conductivity of the diamond-type structure predicted using the SW potential, Tersoff potential, and current MLP, and by DFT calculations.
The phonon dispersion curves and the lattice thermal conductivity predicted using the current MLP are consistent with those predicted by DFT calculations.

\bibliography{SF-AS}
\end{document}